\documentclass[runningheads]{llncs}
\usepackage{graphicx}
\usepackage{textcomp}
\usepackage{gensymb}
\usepackage{hyperref}
\usepackage{multirow}
\usepackage{makecell}
\usepackage[table,xcdraw]{xcolor}
\begin{document}
\title{An anatomically-informed 3D CNN for brain aneurysm classification with weak labels}

%
\author{Tommaso Di Noto\inst{1}\orcidID{0000-0002-5161-055X} \and
Guillaume Marie\inst{1}\orcidID{0000-0002-2447-1056} \and
Sébastien Tourbier\inst{1}\orcidID{0000-0002-4441-899X} \and
Yasser Alemán-Gómez\inst{1,2}\orcidID{0000-0001-6067-8639} \and
Guillaume Saliou\inst{1}\orcidID{0000-0003-3832-7976} \and
Meritxell Bach Cuadra\inst{1,3,4}\orcidID{0000-0003-2730-4285} \and
Patric Hagmann\inst{1}\orcidID{0000-0002-2854-6561} \and
Jonas Richiardi\inst{1}\orcidID{0000-0002-6975-5634}} 

\authorrunning{Di Noto et al.}
%
\institute{Department of Radiology, Lausanne University Hospital and University of Lausanne, Lausanne, Switzerland \and
Center for Psychiatric Neuroscience, Department of Psychiatry, Lausanne University Hospital and University of Lausanne, Lausanne, Switzerland \and
Medical Image Analysis Laboratory, Center for Biomedical Imaging, Lausanne, Switzerland \and
Signal Processing Laboratory (LTS5), Ecole Polytechnique Fédérale de Lausanne, Lausanne, Switzerland}
\maketitle              
\begin{abstract}
A commonly adopted approach to carry out detection tasks in medical imaging is to rely on an initial segmentation. However, this approach strongly depends on voxel-wise annotations which are repetitive and time-consuming to draw for medical experts. An interesting alternative to voxel-wise masks are so-called ``weak" labels: these can either be coarse or oversized annotations that are less precise, but noticeably faster to create. In this work, we address the task of brain aneurysm detection as a patch-wise binary classification with weak labels, in contrast to related studies that rather use supervised segmentation methods and voxel-wise delineations. Our approach comes with the non-trivial challenge of the data set creation: as for most focal diseases, anomalous patches (with aneurysm) are outnumbered by those showing no anomaly, and the two classes usually have different spatial distributions. To tackle this frequent scenario of inherently imbalanced, spatially skewed data sets, we propose a novel, anatomically-driven approach by using a multi-scale and multi-input 3D Convolutional Neural Network (CNN). We apply our model to 214 subjects (83 patients, 131 controls) who underwent Time-Of-Flight Magnetic Resonance Angiography (TOF-MRA) and presented a total of 111 unruptured cerebral aneurysms. We compare two strategies for negative patch sampling that have an increasing level of difficulty for the network and we show how this choice can strongly affect the results. To assess whether the added spatial information helps improving performances, we compare our anatomically-informed CNN with a baseline, spatially-agnostic CNN. When considering the more realistic and challenging scenario including vessel-like negative patches, the former model attains the highest classification results (accuracy$\simeq$95\%, AUROC$\simeq$0.95, AUPR$\simeq$0.71), thus outperforming the baseline. 

\end{abstract}
\section{Introduction}
Cerebral aneurysms (CA) are abnormal focal dilatations in brain arteries caused by a weakness in the blood vessel wall.
The overall population prevalence of CA ranges from 5\% to 8\%~\cite{Rinkel_et_al} and CA rupture is the predominant cause of nontraumatic subarachnoid hemorrhages (SAH)~\cite{Jaja_et_al}. The mortality rate of aneurysmal SAH is around 40\% and only half of post-SAH patients return to independent life~\cite{Frosen_et_al,Xu_et_al}. Considering that the workload of radiologists is steadily increasing~\cite{Rao_et_al,McDonald_et_al} and the detection of CAs is deemed a non-trivial task (especially for small aneurysms)~\cite{Nakao_et_al}, the development of an automatic tool able to detect aneurysms before they become symptomatic would be highly beneficial, both to reduce false negative cases, and to speed up the daily workflow in radiology departments.

Nowadays, non-enhanced Time-Of-Flight Magnetic Resonance Angiography (TOF-MRA) is routinely used for CA detection because of its high sensitivity ($\simeq$ 95\%) and pooled specificity of 89\%~\cite{Chen_et_al}. Also, it has the advantage of being non-invasive and without radiation exposure, as opposed to Digital Subtraction Angiography (DSA) or Computed Tomography Angiography (CTA).

In the last few years, several medical imaging tasks such as classification, detection and segmentation have been profoundly revolutionized by the application of deep learning (DL) algorithms~\cite{Syeda_et_al} which have shown a noteworthy potential. However, DL has to deal with the recurrent challenge of limited availability of (labelled) training examples, for building predictive algorithms that do not suffer from overfitting~\cite{Shen_et_al}. This is especially true in radiology where the voxel-wise manual annotation of medical images is commonly considered a tedious and time-consuming task~\cite{Razzak_et_al} which often takes away precious time from experts.

The task of automated brain aneurysm detection with DL algorithms has already been addressed by several research groups. For instance,~\cite{Ueda_et_al} used 2D patches and a ResNet-like model to detect aneurysms from TOF-MRA. Similarly, 2D Maximum Intensity Projection (MIP) patches with Convolutional Neural Network (CNN) have been proposed by~\cite{Nakao_et_al,Stember_et_al}. In~\cite{Dai_et_al}, 2D nearby projection (NP) images extracted from 3D CTA are fed as input to a Region-CNN (R-CNN) to detect aneurysms. Other works rather use 3D patches to perform aneurysm detection either in MRA or CTA imaging~\cite{Sichtermann_et_al,Park_et_al}. Though many of these works present encouraging results for the development of a Computer-Assisted Diagnosis (CAD) system for aneurysm detection, most of them~\cite{Nakao_et_al,Stember_et_al,Sichtermann_et_al,Park_et_al} build their supervised models starting from voxel-wise manual annotations. From these annotations, they either carry out plain aneurysm segmentation~\cite{Park_et_al}, or they first perform a segmentation and then refine it with post-processing steps~\cite{Nakao_et_al,Stember_et_al,Sichtermann_et_al}, in order to obtain detection bounding boxes.

Differently from previous approaches, our work investigates the task of brain aneurysm classification exploiting ``weak" labels. In our application, these correspond to manual annotations which are not drawn with voxel-wise precision, but rather consist of spheres enclosing the aneurysms which are faster to create for the expert with respect to a slice-by-slice labelling. The concept of ``lazy" or ``weak" labels has already been used in previous works, in particular for segmentation~\cite{Ke_et_al,Ezhov_et_al}, or cell type concentration prediction~\cite{Abousamra_et_al} where full labelling would be infeasible. 
However, while~\cite{Ke_et_al,Abousamra_et_al} exploited under-labelled data, we use over-labelled (more labelled voxels than actual true positives) data to perform aneurysm classification.

The goal of this study is three-fold: first, we assess the capability of a custom CNN to distinguish 3D TOF-MRA patches positive/negative for aneurysms, using weak labels. Second, we show the substantial impact that negative sampling can have on classification performances. Lastly, we propose an anatomically-driven solution to mitigate the problem of negative sampling for our dataset and for similar medical imaging tasks.

\section{Materials and Methods}
\subsection{Data set}
A retrospective cohort of 214 subjects who underwent clinically-indicated TOF-MRA between 2010 and 2012 was used. Out of these 214 subjects, 83  had one (or more) aneurysm(s), while 131 did not present any. For the former group, patients with one or more unruptured intracranial aneurysms were included, while patients with treated and ruptured aneurysms were excluded.  Different aneurysms of the same patient were treated as independent, but most patients (81\%) had only one aneurysm. Similarly, for patients with multiple sessions, we treated each session independently. The overall number of aneurysms included in the study is 111 and their anatomical location distribution is shown in Table~\ref{aneur_spatial_distribution}. A 3D gradient recalled echo sequence with Partial Fourier technique was used for all subjects (see MR acquisition parameter details in Table \ref{MR_parameters}). Aneurysms were annotated by one radiologist with 4 years of experience in neuroimaging. The \href{http://ric.uthscsa.edu/mango/}{Mango software} was used to create the aforementioned weak labels which correspond to spheres that enclose the whole aneurysm, regardless of the shape (i.e. saccular, fusiform or multilocular). All TOF-MRA subjects included in the study were double checked by a senior neuroradiologist with over 14 years of experience, in order to exclude potential false positives or false negatives that might have been present in the original medical reports. The data set was organized according to the Brain Imaging Data Structure (BIDS) standard~\cite{Gorgolewski_et_al}.

\begin{table}[ht]
\centering
\caption{Spatial distribution of aneurysms. MCA = Middle Cerebral Artery, ACOM = Anterior Communicating Artery, PCOM = Posterior Communicating Artery.}\label{aneur_spatial_distribution}
\begin{tabular}{|
>{\columncolor[HTML]{FFFFFF}}c |
>{\columncolor[HTML]{FFFFFF}}c |
>{\columncolor[HTML]{FFFFFF}}c |}
\hline
                                  & \textbf{Count}            & \textbf{\%} \\ \hline
\textbf{MCA}                      & 22                        & 19.8        \\ \hline
\textbf{ACOM}                     & {\color[HTML]{000000} 20} & 18.0        \\ \hline
\textbf{Intradural carotid other} & 13                        & 11.7        \\ \hline
\textbf{Carotid extra}            & 13                        & 11.7        \\ \hline
\textbf{MC other}                 & 8                         & 7.2         \\ \hline
\textbf{Carotid tip}              & 8                         & 7.2         \\ \hline
\textbf{Pericallosal}             & 8                         & 7.2         \\ \hline
\textbf{PCOM}                     & 7                         & 6.3         \\ \hline
\textbf{Basilar tip}              & 5                         & 4.5         \\ \hline
\textbf{Ophthalmic}               & 4                         & 3.6         \\ \hline
\textbf{Post other}               & 3                         & 2.7         \\ \hline
\end{tabular}
\end{table}

\begin{table}
\centering
\caption{MR acquisition parameters of TOF-MRA scans used for the study population.}\label{MR_parameters}
\scriptsize
\begin{tabular}{|c|c|c|c|c|c|c|c|c|}
\hline
\textbf{\# scans} &  \textbf{Vendor} & \textbf{Model} & \multicolumn{1}{c|}{\textbf{\begin{tabular}[c]{@{}c@{}}Field\\ Strength\\ {[}T{]}\end{tabular}}} & \textbf{\begin{tabular}[c]{@{}c@{}}TR\\ {[}ms{]}\end{tabular}} & \textbf{\begin{tabular}[c]{@{}c@{}}TE\\ {[}ms{]}\end{tabular}} & \textbf{\begin{tabular}[c]{@{}c@{}}Pixel\\  spacing\\ {[}mm\(^2\){]}\end{tabular}} & \multicolumn{1}{c|}{\textbf{\begin{tabular}[c]{@{}c@{}}Slice \\ Thickness\\ {[}mm{]}\end{tabular}}} & \multicolumn{1}{c|}{\textbf{\begin{tabular}[c]{@{}c@{}}Slice\\ Gap\\ {[}mm{]}\end{tabular}}}\\
\hline
81 &  Philips & Intera & 3.0 & 18.3 & 3.40 & 0.41x0.41 & 1.1 & 0.55\\
\hline
10 &  \begin{tabular}[c]{@{}c@{}}Siemens\\ Healthineers\end{tabular} & Aera & 1.5 & 24.0 & 7.0 & 0.35x0.35 & 0.5 & 0.09\\
\hline
21 &  \begin{tabular}[c]{@{}c@{}}Siemens\\ Healthineers\end{tabular} & Skyra & 3.0 & 21.0 & 3.43 & 0.27x0.27 & 0.5 & 0.08\\
\hline
35 &  \begin{tabular}[c]{@{}c@{}}Siemens\\ Healthineers\end{tabular} & Symphony & 1.5 & 39.0 & 5.02 & 0.39x0.39 & 1 & 0.25\\
\hline
28 &  \begin{tabular}[c]{@{}c@{}}Siemens\\ Healthineers\end{tabular} & TrioTim & 3.0 & 23.0 & 4.18 & 0.46x0.46 & 0.69 & 0.14\\
\hline
61 & \begin{tabular}[c]{@{}c@{}}Siemens\\ Healthineers\end{tabular} & Verio & 3.0 & 22.0 & 3.95 & 0.46x0.46 & 0.70 & 0.13\\
\hline
\end{tabular}
\end{table}

\subsection{Image processing}
Two preprocessing steps were carried out for each subject. First, we performed skull-stripping with the Brain Extraction Tool~\cite{Smith_et_al} to remove regions such as the skull or the eyes.
Second, a probabilistic vessel atlas built from multi-center MRA data sets~\cite{Mouches_et_al} was co-registered to each patient's TOF-MRA using the Advanced Neuroimaging Tools (ANTS)~\cite{Avants_et_al}. Specifically, we first registered the probabilistic atlas to the T1-weighted anatomical scan of each patient through a symmetric diffeomorphic registration. Second, we registered the obtained warped volume to the TOF subject space through an affine registration. The registered atlas was used only to provide prior information about vessel locations for the patch sampling strategy (see \ref{sampling} below). As for most of the previously mentioned studies, we adopt a patch-based approach for the classification of aneurysms: we use 3D TOF-MRA patches as input samples to our network, rather than the entire volumes.

\subsection{An anatomically-informed 3D-CNN}
\begin{figure}[ht]
\centering
\includegraphics[width=0.8\textwidth]{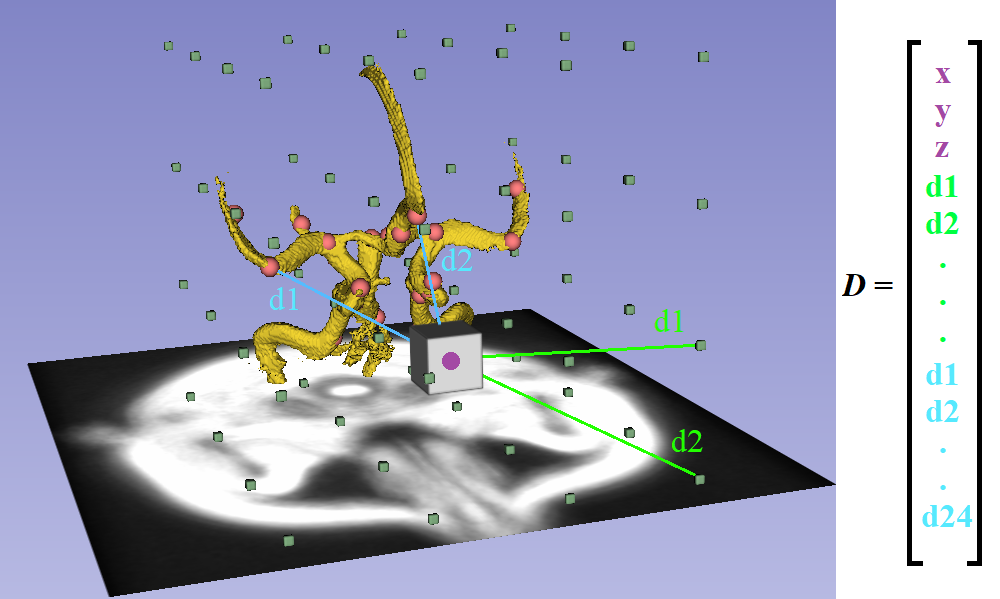}
\caption{Computation of spatial features: for each input patch, we extract a feature vector $\mathbf{D}$ which is composed of the [x,y,z] coordinates of the patch center (in purple), the distances from this center to the points of a uniform grid (light green) and the distances to some landmarks (red dots) recurrent for aneurysms (light blue). The main brain arteries are segmented from the vessel atlas and are depicted in yellow.} \label{spatial_features}
\end{figure}

The task of aneurysm classification is extremely spatially constrained, since not only aneurysms solely occur in arteries, but they also occur in precise locations of the vasculature that have higher probability than others. Inspired by previous works in neuroimaging~\cite{Ghafoorian_et_al,Ganaye_et_al}, we decided to include this strong anatomical prior into our model. This was achieved in two ways: first, we designed a two-channel CNN which analyzes patches of the input volume at different spatial scales, in order to provide anatomical context on the vascular tree surrounding the patch of interest. Second, we computed for each input sample a numerical vector $\mathbf{D}$ containing tailored spatial features which is integrated in the fully connected layers of the network; namely, the vector $\mathbf{D}$ includes the [x,y,z] coordinates of the center of the input patch in MNI space, and the Euclidean distances from this center to several coordinate and landmark points (also in MNI space). The coordinate points belong to a cubic (6x6x6), uniformly-sampled grid superimposed on the TOF-MRA volume. The landmark points correspond to 24 arterial locations where aneurysms are most recurrent; these were selected basing on the literature~\cite{Brown_et_al} and on our data. The final dimension of $\mathbf{D}$ is 243. A 3D visual representation of these distances and of the creation of $\mathbf{D}$ are provided in Figure~\ref{spatial_features}.

\begin{figure}[ht]
\centering
\includegraphics[width=\textwidth]{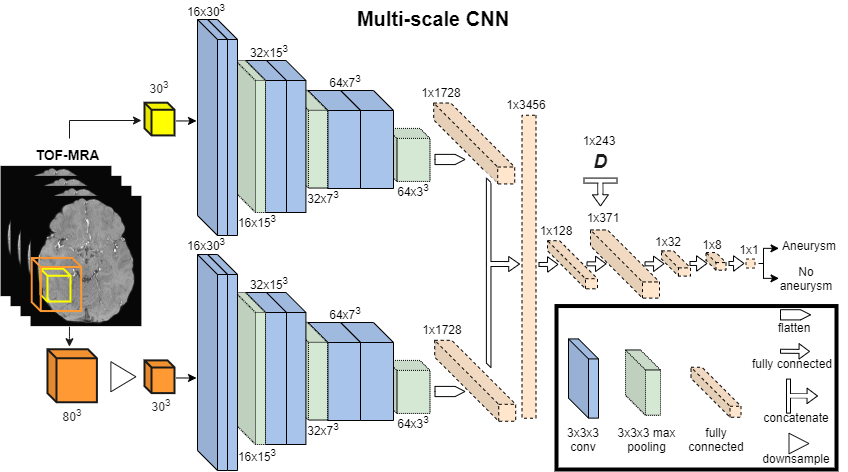}
\caption{CNN architecture: features are extracted in parallel from small-scale and large-scale TOF-MRA patches through a stack of convolutional layers. Then, they are merged into a single fully connected layer. Later, the spatial information vector $\mathbf{D}$ is concatenated to a fully-connected layer.} \label{CNN_structure}
\end{figure}

\textbf{Network architecture -} We designed a custom CNN with building blocks inspired by the VGG-16 network~\cite{Simonyan_et_al}. Figure~\ref{CNN_structure} illustrates in detail the structure of our CNN. As already proposed in~\cite{Huang_et_al,Nie_et_al}, we performed late fusion of the features extracted from the input channels. Essentially, the same stack of convolutional layers is applied in parallel both to the small-scale and large-scale TOF-MRA patches. Then, feature vectors are merged, passed to a stack of fully connected layers, and further concatenated with vector $\mathbf{D}$. The rationale behind the multi-scale approach is that the convolutions over the small-scale patches produce aneurysm-specific features, whereas the large-scale patches provide context/spatial descriptors. Since most of the aneurysms in our dataset (92\%) had an equivalent diameter smaller than 30 voxels, we decided to fix the side of the small-scale input patches to 30. Instead, a side of 80 was set for the large-scale patches in order to include even the largest aneurysm (equivalent diameter = 58 voxels) and some context around it. All patches were standardized to have mean 0 and variance 1 before being fed to the CNN as suggested by~\cite{Goodfellow_Bengio_et_al}. The standardization was also performed to mitigate intensity differences which are inherently present across different patients and scanners~\cite{Zhuge_et_al}. A kernel size of 3x3x3 was used in all convolutional layers, with padding and a stride=1 in all directions. We applied the Rectified Linear Unit (ReLU) activation function for all layers, except for the last fully connected layer which is followed by a sigmoid function. To fit the model, the Adam optimization algorithm~\cite{Kingma_et_al} was applied with variable learning rate, together with the binary cross-entropy loss. Moreover, we used the Xavier initialization~\cite{Glorot_et_al} for all the layers of the CNN. Biases were initialized to 0 and a batch size of 4 was chosen. The final output of the CNN is simply the class probability $p$ of the input sample: positive (patch with aneurysm) if $p>0.5$ or negative (without aneurysm) if $p<0.5$.

To elucidate whether the injection of anatomical and spatial information into the model can improve classification results, we compare two distinct architectures: the \textbf{anatomically-informed 3D-CNN} illustrated in Figure~\ref{CNN_structure} and a \textbf{baseline} CNN which is identical to the previous one, but has one input channel (i.e. only the small-scale TOF-MRA) and no spatial features (i.e. no vector $\mathbf{D}$). The former has 884,529 trainable parameters, while the latter has 448,417.

\subsection{Patch sampling strategy}
\label{sampling}
In addition to the comparison between models (anatomically-informed and baseline), we also investigated the influence that negative sampling can have on classification performances. Indeed, while for the minority class (i.e. patches with aneurysm) the sampling is restricted by the availability of positive cases, extraction of negative samples (majority-class sampling) necessarily entails the choice of one (or more) extraction criteria. Therefore, we chose two different sampling criteria for extracting negative patches:

\textbf{1) Random sampling:} negative samples were extracted randomly within the skull-stripped brain, without overlapping with the positive patches.

\textbf{2) Intensity-matched sampling:} we imposed intensity constraints for the extraction basing on the co-registered vessel atlas. More specifically, with an iterative search, we only extract the negative sample when the corresponding (i.e. same center coordinates) vessel patch has both a local (patch-wise) and global (volume-wise) brightness that are higher than some specific thresholds. These thresholds, in turn, were chosen empirically according to the local and global brightness of all positive patches in the vessel atlas. First, this sampling strategy avoids extracting patches that are too dark with respect to positive ones. Second, it allows us to extract patches which always include part of the vasculature. Needless to say, this sampling creates both more realistic and more difficult negative samples for the CNN.

Instead, positive patches were extracted around the aneurysms in a non-centered fashion, but ensuring that the aneurysm mask was always completely included in the small-scale patch. As last step, regardless of the sampling strategy, we combined the samples (negative and positive) of all subjects into a unique dataset that was fed as input to the CNN. We decided to extract 8 negative samples per subject. This led to a final dataset composed of 1808 negative and 111 positive samples (ratio $\simeq 1:16$). During training, a series of data augmentation techniques were applied on positive patches: namely, rotations (90\degree,180\degree,270\degree), horizontal flip, elastic deformation and contrast adjustment. Training and evaluation were performed with Tensorflow 2.0 and a GeForce RTX 2080TI GPU.

\subsection{Evaluation approach}
We evaluated all different scenarios through a nested stratified Cross Validation (CV), with 5 external folds and 3 internal folds. This ensured that the patches in every test fold were always unseen samples with respect to the training set and to the chosen hyperparameters. The only hyperparameter that was tuned in the internal CV is the learning rate: 0.00001, 0.0001 and 0.001 were tested. All other hyperparameters were fixed, so as not to excessively lengthen training time. We set a dropout rate of 0.2, a sample weight factor of 3 to give more importance to the minority class and we trained the CNN for 50 epochs with an early stopping condition on the validation/test set. To statistically compare classification results, Wilcoxon signed-rank tests were performed~\cite{Wilcoxon_et_al}. For simplicity, the tests only accounted for the area under the PR (AUPR) curve of the classifiers, since this metric is particularly suited when working with imbalanced data sets~\cite{Davis_et_al}.
A significance threshold level $\alpha=0.05$ was set for comparing P values.
First, we saved the best hyperparameters for each experiment. Then, we re-ran the training/test of the CNN 10 times. For each of the 10 realizations we always changed the patient order and ensured that the two models (baseline and anatomically-informed) were evaluated against the exact same samples. 

\section{Results}
Overall, \textbf{4 experiments} were carried out: the two networks (anatomically-informed and baseline) were evaluated first against the dataset with random negative samples and then against the dataset with intensity-matched negative samples. Classification results of the four experiments are reported in Table~\ref{class_results_table}. Training the model took about 2 hours for the baseline model and 3 hours for the anatomically-informed one. The most frequent learning rate across the external test folds was 0.0001.
The Wilcoxon tests performed on the AUPRs distributions highlighted two main findings: first, both for the baseline and the anatomically-informed model, AUPRs were statistically higher when random negative sampling was used with respect to the intensity-matched sampling ($P = 0.01$). This proves how one task is evidently easier than the other. Second, when comparing the baseline model and the anatomically-informed model against the intensity-matched dataset (difficult scenario), AUPR distributions were again significantly different ($P = 0.01$), suggesting that the proposed anatomically-informed CNN indeed outperforms the baseline.

\begin{table}[ht]
\label{class_results_table}
\centering
\caption{Classification results of baseline and anatomically-informed models both with random negative patches and with intensity-matched (IM) ones. Acc=accuracy, Sens=sensitivity, Spec=specificity, PPV=positive predictive value, NPV=negative predictive value, AUC=Area Under ROC Curve, AUPR=Area Under PR curve.}
\scriptsize
\begin{tabular}{|c|c|c|c|c|c|c|c|c|}
\hline
\textbf{Network}                                                                 & \textbf{\begin{tabular}[c]{@{}c@{}}Negative\\ Samples\end{tabular}} & \textbf{Acc (\%)} & \textbf{Sens (\%)} & \textbf{Spec (\%)} & \textbf{PPV (\%)} & \textbf{NPV (\%)} & \textbf{AUC} & \textbf{AUPR} \\ \hline
\multirow{2}{*}{Baseline}                                                        & \textbf{Random}                                                     & 95.4              & 84.7               & 96.2               & 59.7              & 99.0              & .961         & .779          \\ \cline{2-9} 
                                                                                 & \textbf{IM}                                                         & 93.2              & 77.5               & 94.1               & 45.8              & 98.6              & .949         & .608          \\ \Xhline{2\arrayrulewidth}
\multirow{2}{*}{\begin{tabular}[c]{@{}c@{}}Anatomically\\ Informed\end{tabular}} & \textbf{Random}                                                     & 97.3              & 91.9               & 97.7               & 72.6              & 99.5              & .979         & .875          \\ \cline{2-9} 
                                                                                 & \textbf{IM}                                                         & 94.7              & 77.5               & 95.7               & 53.6              & 98.6              & .946         & .714          \\ \hline
\end{tabular}
\end{table}

\section{Discussion}
This work presented an alternative approach for performing cerebral aneurysm detection when voxel-wise annotations are not available. To this end, we proposed a binary classification method, making use of weak labels enclosing the aneurysms of interest. In addition, we shed light over the recurrent problem of negative sampling in imbalanced and spatially-skewed data sets, showing how this step can dramatically alter final results. Lastly, we devised a tailored CNN able to mitigate the negative sampling problem, by incorporating spatial anatomical information. This CNN was able to outperform its baseline counterpart despite the small sample size and having about twice the number of parameters. We believe this general principle is applicable to several other brain diseases with sparse spatial extent. 

Our work is limited by the relatively high number of false positive cases even for the anatomically-informed CNN (see low PPV in Table \ref{class_results_table}). In addition, a separate analysis should be performed to understand whether the added distances of vector $\mathbf{D}$ are indeed helpful: these might be redundant with respect to the [x,y,z] center coordinates of the patches, which could already be informative enough. Though the presented patch-wise analysis is useful to gain insights on the network performances, it cannot be easily exploited in a clinical scenario. Thus, future work will aim at shifting towards a patient-wise analysis. Lastly, we acknowledge that the dataset size is still limited and it should be increased.

%
%
%
%

\end{document}